\documentclass[journal abbreviation, manuscript]{copernicus}
\nolinenumbers

\usepackage{graphicx}
\usepackage{array}
\usepackage{ chemformula }

\begin{document}

\title{Similarity-Based Analysis of Atmospheric Organic Compounds for Machine Learning Applications}


\Author[1]{Hilda}{Sandström}
\Author[1,2,3][patrick.rinke@aalto.fi]{Patrick}{Rinke} 

\affil[1]{Department of Applied Physics\\
Aalto University\\
P.O. Box 11000\\
FI-00076 Aalto, Espoo, Finland\\}
\affil[2]{Physics Department, TUM School of Natural Sciences, Technical University of Munich, Garching, Germany}
\affil[3]{Munich Data Science Institute, Technical University of Munich, Garching, Germany}
\runningtitle{Similarity-Based Analysis of Atmospheric Organic Compounds for Machine Learning Applications}

\runningauthor{H. Sandstr\"om, Patrick Rinke}

\received{}
\pubdiscuss{} 
\revised{}
\accepted{}
\published{}

\firstpage{1}

\maketitle
\begin{abstract}
The formation of aerosol particles in the atmosphere impacts air quality and climate change, but many of the organic molecules involved remain unknown. Machine learning could aid in identifying these compounds through accelerated analysis of molecular properties and detection characteristics. However, such progress is hindered by the current lack of curated datasets for atmospheric molecules and their associated properties. To tackle this challenge, we propose a similarity analysis that connects atmospheric compounds to existing large molecular datasets used for machine learning development. We find a small overlap between atmospheric and non-atmospheric molecules using standard molecular representations in machine learning applications. The identified out-of-domain character of atmospheric compounds is related to their distinct functional groups and atomic composition. Our investigation underscores the need for collaborative efforts to gather and share more molecular-level atmospheric chemistry data. The presented similarity based analysis can be used for future dataset curation for machine learning development in the atmospheric sciences.
\end{abstract}

\introduction  
Aerosol particles influence our climate by sunlight reflection and absorption, and by serving as nuclei for cloud condensation \citep{IPCC2022}. Beyond climate impact, aerosol particles affect air quality, causing adverse effects on human health \citep{Pozzer2023, Khomenko2021, Lelieveld2020}.  However,  the underlying molecular-level processes involving organic molecules remain poorly understood, due to the vast number of organic compounds participating in atmospheric chemistry. This existing gap in knowledge hampers a comprehensive understanding of particle formation and growth in different environments \citep{IPCC2021,Elm2020}. In this paper, we take an initial step to evaluate the potential of filling this void using machine learning. We propose a molecular similarity-based analysis to measure the overlap between atmospheric compounds and common molecular datasets used in machine learning development. By doing so, we can provide a tool to tailor machine learning models for studies of aerosol particle formation and the effects human-based activities, like industry and agriculture, on the formation process. Ultimately, such insights can lead to more informed decisions regarding air quality and climate change mitigation.

Organic aerosol particle formation is affected by atmospheric composition and molecular emissions into the atmosphere.  Emitted molecules can transform in reactions initiated by sunlight, to a diverse array of compounds with numerous functional groups \citep{Bianchi2019}. These reactions are estimated to produce between hundreds of thousands to millions of atmospherically relevant molecules \citep{Goldstein2007, Noziere2015}. Out of this plethora, an unknown number can form or grow aerosol particles by interacting with inorganic emissions \citep{Schobesberger2013, Riccobono2014, Ehn2014}, or by themselves \citep{Kirkby2016}.  Details of aerosol particle formation can be uncovered through identification of relevant atmospheric reactions (e.g., \citep{Perakyla2020, Iyer2023}),  aerosol forming compounds (e.g. \citep{Franklin2022,Worton2017,Hamilton2004, Thoma2022}) and cluster formation steps \citep{Elm2020}. 

Mapping  aerosol particle formation experimentally is challenging due to the sheer number of potentially relevant compounds. Moreover, spectrometry based compound identification with, e.g., mass spectrometers, is hindered by the absence of curated reference spectra for atmospheric molecules \citep{Franklin2022, Worton2017, Hamilton2004}. The study of particle growth is another experimental challenge  due to the wide range of size scales involved. Neither aerosol mass spectrometry nor atmospheric pressure chemical ionization  mass spectrometry alone can be used to study the entire particle growth process \citep{Elm2020}.  Thus, with a few exceptions (e.g., \citep{Franklin2022, Worton2017, Hamilton2004, Sander2015}),  curated structure-annotated molecular datasets from experiments are lacking. 

In adjacent chemical disciplines such as metabolomics, these curated molecular datasets play a crucial role for chemical analysis. For example, they assist in compound identification either directly \citep{Kind2013,Sud2007,mzCloud,Sawada2012,Wissenbach2011_dev,Wissenbach2011_drugs,Montenegro2020,Taguchi2010,wileyTandem,Hummel2013,Watanabe2000,Wiley12,Wang2016,Wishart2022,Wallace2023,Weber2012, MassBankEu22,MONA23} or through the development of machine learning-based identifcation tools \citep{Heinonen2012,Durkop15,Brouard2016, Nguyen2018,Nguyen2019}. These datasets also form the foundation of data-driven analysis platforms e.g., \citep{Nothias2020}. Moreover, curated datasets contribute to the construction of machine learning models for quantitative structure-activity relationships, facilitating large-scale screening of molecular properties for specific reactions or applications  \citep{Kulik2022}. Thus, in order to  reach the full potential of data-driven methods we need such datasets. Currently, in atmospheric science, computational techniques are bridging the gap to what can be experimentally observed for atmospheric compounds.

\begin{figure}[t]
\includegraphics[]{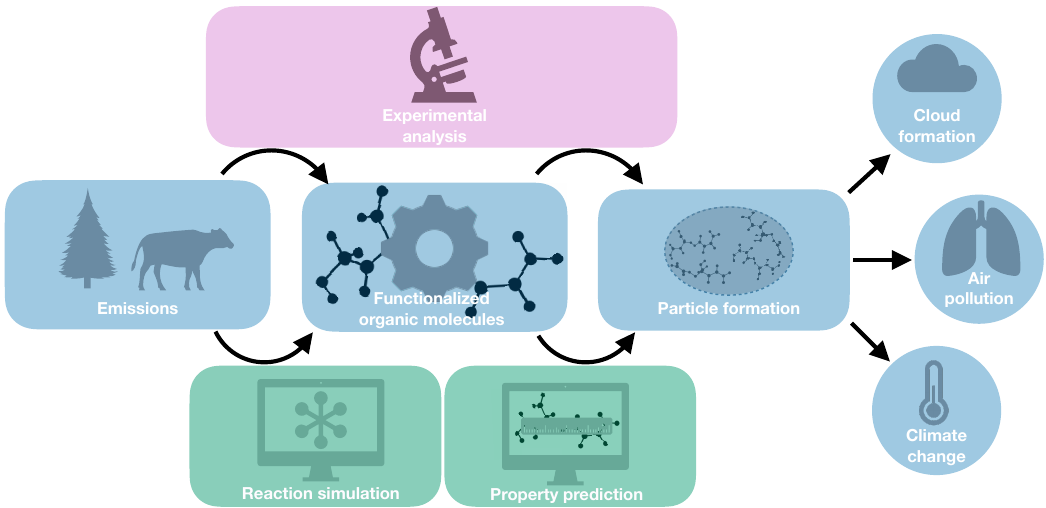}
\caption{Molecular emissions react in the atmosphere to form a diverse array of compounds, contributing to aerosol particle formation. However, identifying these compounds remains challenging. Although experiments and field studies can detect certain compounds in the atmosphere and within the aerosols, they lack the ability to resolve the identity of a majority of these compounds. Computational techniques, like reaction mechanistic simulations and property predictions, aid in describing atmospheric reaction products and their impact on particulate matter. Enhancing our understanding of these molecular processes will illuminate the effects of human emissions on cloud formation, air quality, and climate. Data-driven methods could help advance and accelerate both the displayed experimental and computational workflow.}
\label{fig:atmchem}
\end{figure}

Computational simulations and predictive modeling offer an alternative approach to studying molecular-level atmospheric chemistry (Figure \ref{fig:atmchem}).  Reaction models, such as Gecko-A \citep{Aumont2005} or the Master Chemical Mechanism  (MCM, http://mcm.
leeds.ac.uk/MCM)\citep{Jenkin1997, Saunders2003}, can be used to propose likely atmospheric reaction products based on a set of  precursor molecules, reactions and conditions. With such model simulations, atmospheric molecular datasets like Gecko \citep{vanwertz2021} and Wang \citep{Wang2017} have been generated. The Wang dataset \citep{Wang2017} was constructed using MCM \citep{Jenkin1997, Saunders2003} to simulate the atmospheric degradation of 143 atmospheric compounds by photolysis and reactions with \ch{OH}, \ch{NO3} and \ch{O3}. Similarly, the Gecko dataset \citep{vanwertz2021} was generated by simulating the gas phase oxidation of three important atmospheric compounds: toluene, $\alpha$ - pinene and decane, using the Gecko-A code \citep{Aumont2005, Lannuque2018}. Both the Wang and Gecko datasets have been used to predict physicochemical properties such as saturation vapor pressures and partition coefficients \citep{Wang2017,Lumiaro2021,Besel2023,Besel2024}.  In addition, computational simulations of particle formation have resulted in the Clusteromics datasets containing common acid-base clusters and their associated thermodynamic and kinetic properties \citep{Elm2019, Elm2021C1, Elm2021, elm2022,Knattrup2022}. Thus, simulations and property prediction can be used to propose important candidate compounds in organic aerosol formation processes (Figure \ref{fig:atmchem}). 

In recent years, machine learning methods have shown promise for accelerating  traditional computational and experimental atmospheric chemistry research  \citep{sandstroem2024,Franklin2022,Besel2023,Besel2024,Berkemeier2023,Kubecka2023-curr,Knattrup2023,Kruger2022,Hyttinen2022,Lumiaro2021}. Yet, practical applications of data-driven methods in atmospheric chemistry are still hindered by the aforementioned scarcity of curated experimental datasets. This raises questions about how machine learning advancements in atmospheric chemistry can leverage molecular datasets and models from computational simulations or other chemical disciplines. While this approach is currently especially important in atmospheric chemistry, it also mirrors similar efforts of data augmentation in other fields.

Here, our goal is to assess how closely atmospheric molecular data aligns with comprehensive curated datasets from other chemical domains. We assess the impact of the current data gap in atmospheric chemistry on the progress of data-driven methods in the field. We also address the potential for using datasets and models developed in other tangential chemical domains (e.g. metabolomics, drug design or environmental chemistry) for transfer learning or data augmentation in atmospheric chemistry.

\begin{figure}[H]
\includegraphics{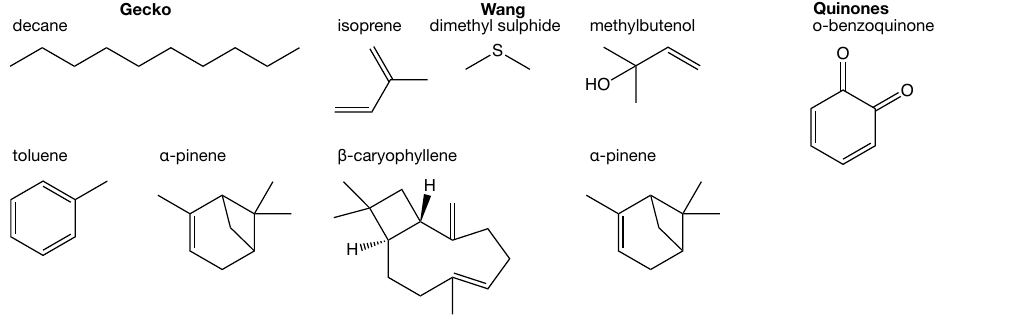}
\caption{The three atmospheric datasets used in the comparison of this paper are: Wang \citep{Wang2017}, Gecko \citep{vanwertz2021}, and Quinones \citep{Tabor2019,Kruger2022}. The Gecko and Wang datasets contain simulated reaction products starting from a set of precursors (exemplified in graphic), and the Quinone dataset  contains compounds from a class of oxidation products derived from aromatics called quinones.}
\label{fig:atm_precur}
\end{figure}

In our analysis, we represent atmospheric compounds by the above mentioned Wang and Gecko datasets. In addition, we include a third  atmospheric dataset composed of quinones, organic molecules that result from oxidation of aromatic compounds \citep{Tabor2019,Kruger2022}. Example molecular precursors to Wang and Gecko, as well as an example quinone compound are shown in Figure \ref{fig:atm_precur}.

We explore the similarity of our atmospheric compound domain to four molecular datasets used either for molecular property prediction (QM9 and nablaDFT) or compound identification by mass spectrometry (MassBank Europe and MassBank of North America, MONA). Both application areas are relevant to machine learning in molecular atmospheric research  \citep{sandstroem2024, Franklin2022, Worton2017, Hamilton2004, Besel2023,Lumiaro2021,Wang2017}. QM9 \citep{Ramakrishnan2014} is a standard benchmark dataset for machine learning in molecular sciences. The dataset was constructed by selecting molecules with up to nine non-hydrogen atoms (limited to OCNF) from  the GDK-17 database \citep{Ruddigkeit2012}. GDK-17 is a database consisting of 166 billion enumerated molecules consisting of up to 17 CNOS and halogen atoms.  QM9 includes  harmonic frequencies, dipole moments, polarizabilities,  and electronic and thermal energies for the molecular minimum energy conformation. nablaDFT, instead, was  curated from the Molecular Sets (MOSES) dataset \citep{Polykovskiy2020} for the purpose of training models for quantum chemical property prediction (conformational energy and Hamiltonian). On the other hand, the MassBanks provide data pairs of molecular structures and their corresponding mass spectra, and have been used to train and test machine learning models for compound identification based on mass spectra \citep{Heinonen2012, Durkop15, Durkop2019}. The MassBanks primarily contain  molecules with relevance to metabolomics or environmental studies. While MassBank Europe contains purely experimental data, MONA also provides computationally predicted mass spectra. Table \ref{tab:datasets} summarizes the seven atmospheric and non-atmospheric molecular datasets we use in our analysis.  

\begin{table}
\caption{The molecular datasets used in our similarity analysis comparing atmospheric compounds to metabolite and drug compounds. The datasets were downloaded January 8th 2024 in SMILES (Simplified molecular-input line-entry system) format (xyz in the case of QM9). The reported dataset sizes were obtained after data preprocessing, which involved removing unparsable SMILES  representations and eliminating duplicate entries.}
\label{tab:my-table}
\resizebox{\textwidth}{!}{%
\begin{tabular}{|l|l|p{3cm}|p{4cm}|p{1cm}|p{1cm}|l|}\hline
Name                             & Data instances   & Type of compound     & Elements  & Exp. data & Comp. data & Ref. \\ \hline
Gecko    & 166434  & Atmospheric &   C, O, H, N&  N & Y & \citep{vanwertz2021}    \\
Wang     & 3414    & Atmospheric &   C, O, N, H, Cl, S, Br& N & Y & \citep{Wang2017}    \\
Quinones & 69599  & Atmospheric &   C, N, O, H, P&  N & Y & \citep{Tabor2019,Kruger2022}\\
nablaDFT & 1004918 & Druglike    &   Br, C, N, H, O, S, Cl, F& N & Y & \citep{Polykovskiy2020,Khrabrov2022} \\
QM9 & 133885 & Druglike &    O, C, H, N, F& N               & Y                  & \citep{Ruddigkeit2012,Ramakrishnan2014} \\
MONA& 681692 & Majority metabolites and drug molecules &   H, C, O, Cl, Si, I, S, N, Br, F, Na, P, Co, B, K, Fe, Ge, Sn, Cu, Mg, Pd, Al, Ni, Pt, Cr, Au, Se, Zn, Hg, As&  Y                 & Y                  & \citep{MONA}\\
MassBank Europe & 21772 & Small molecules relevant to metabolomics, exposomics and environmental samples   & C, O, H, Cl, Si, I, S, N, B, Br, F, P, As, Ge, Sn, Cu, Na, Pd, Al, Co, Ni, Pt, Se, Zn, Hg, K& Y & N & \citep{MassBankEu22}\\\hline
\end{tabular}
}
\label{tab:datasets}
\end{table}

The paper is organized as follows. We present our molecular similarity analysis method in Section \ref{sec:met}. Section \ref{sec:res} presents the outcomes of our similarity based comparison. In Section  \ref{sec:dis} we discuss our findings and in Section  \ref{sec:out}  we provide an outlook on how our similarity based analysis can be used to guide data curation for model development in atmospheric research.

\section{Methods}\label{sec:met}

\subsection{Molecular similarity} In our similarity-based analysis, we measure the overlap between atmospheric compounds and other sets of molecules using the two similarity metrics t-stochastic neighbor embedding (t-SNE \citep{Maaten2008}, as implemented in scikit-learn v. 1.2 \citep{scikit-learn}), and the Tanimoto similarity index \citep{Tanimoto1958} as implemented in RDKit version 2022.09.3 \citep{Landrum2022}. These metrics utilize a molecular representation in the form of a binary vector (see Molecular descriptors below). Both t-SNE and the Tanimoto similarity index are standard tools to measure chemical diversity (see e.g., \citep{Soleimany2023, Nakamura2022}) for out-of-domain applications and  uncertainty quantification \citep{Moret2023, Hirschfeld2020, Scalia2020, Janet2019, Liu2018, Sheridan2004}. t-SNE is an unsupervised machine learning method that embeds high-dimensional data into lower dimensions while preserving distances from the higher-dimensional space. The low dimensional embedding can be used to draw qualitative conclusions about data structure and similarity.  We tested three different perplexity values of 5, 50 and 100. We pre-process the molecular fingerprints by performing a principal component analysis and select the 50 first components. The t-SNE clusters depend on a perplexity hyperparameter which in brief balances the preservation  of global and local aspects during  projection from high to low-dimensional space. Thereafter, we run the t-SNE clustering with random initialization for a maximum of 5000 iterations. 

By contrast, the Tanimoto index, $S_{A,B}$, offers a quantitative measure of similarity.  $S_{A,B}$ is calculated as the fraction of present features (represented by non-zero bits) that are shared compared to the number of unshared features between molecules A and B, according to
\begin{equation}
S_{A,B} = \frac{\sum_{A \cap B}1}{(\sum_{A \cup B}1-\sum_{A \cap B}1)}. 
\label{eq:tanimoto}
\end{equation}
If the two  molecules A and B share all features, the Tanimoto index equals to one and if they instead share no features, equals to zero. We note that both similarity metrics depend on the choice of molecular representation. Here, we performed the analysis with two types of molecules representations (see Molecular descriptors below). 

We make a statistical analysis of the Tanimoto similarities between pairs of molecules from different datasets, comparing two sets at a time. Initially, we select either the Wang dataset or the Gecko dataset as our reference dataset. Then, we compute the Tanimoto similarity for each molecule in the non-reference dataset against every molecule in the chosen reference dataset. This process yields a distribution of pairwise similarities, illustrating the degree of resemblance between molecules in the non-reference dataset and those in the reference dataset. Additionally, we calculate the self-similarity within the reference dataset by determining the Tanimoto similarity for all pairs of molecules within it. Analyzing the obtained similarity distributions allows us to assess the relationship between the datasets and understand both the inter-dataset similarities and internal similarity of the reference dataset. We interpret our results by introducing a high and low similarity reference values. This choice is motivated by previous studies of Tanimoto similarity \citep{Liu2018,Moret2023}. A similarity of 0.1 or less is considered to indicate no significant molecular similarity \citep{Liu2018}. Moreover,  a nearest neighbor similarity to the training set greater than 0.4 has been shown to indicate improved machine learning model prediction performance \citep{Liu2018} and confidence \citep{Moret2023}.

\subsection{Molecular descriptors} 
\begin{figure}[H]
\includegraphics[]{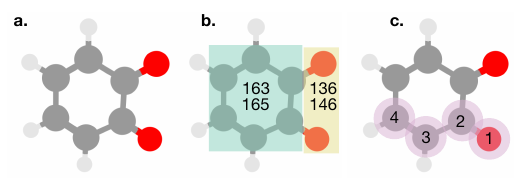}
\caption{Pictoral overview of the two molecular fingerprints used in our analysis. (a) A ball-and-stick representation of o-benzoquinone. (b) The MACCS fingerprint contains 166 keys which correspond to answers to yes-or-no questions regarding the presence of molecular features, such as whether there is a ring and if it is 6-membered (keys 163 and 165), or if there is more than one oxygen, or more than one double bonded oxygen (keys 136 and 146). (c) The topological fingerprint encodes paths in the two-dimensional molecular structure. The panel shows an example path length traversing four atoms. }
\label{fig:descr}
\end{figure}

As mentioned above, we perform our similarity analysis with two different molecular representations as implemented in RDKit: the RDKit topological fingerprint \citep{Landrum2022} and the Molecular ACCess System (MACCS) fingerprint \citep{Accelrys2011}. The MACCS fingerprint consists of 166 keys (RDKit’s version has 167 keys as key 0 is an unused dummy key) which represent different molecular features (e.g., number of rings or atoms of a certain element, Figure \ref{fig:descr}b).The topological fingerprint is based on enumeration of paths in the 2D-molecular structure (Figure \ref{fig:descr}c). We used default hyperparameter values  for the topological fingerprint (a maximum path length of seven, two bits per hash, and a fingerprint length of 2048 bits).  A wide variety of fingerprints and molecular representations have been developed in cheminformatics and more recently in chemistry, physics and materials science \citep{Himanen2020,Langer2022}. In this paper, we limit ourselves to the topological and the MACCS fingerprints out of practicality and relevance.  Both fingerprints have been used in atmospheric chemistry machine learning applications \citep{Lumiaro2021, Besel2023, Besel2024} and are therefore pertinent for our comparison. 

We performed a molecular structure analysis in RDKit, and the functional group analysis using the APRL-substructure search program \citep{Ruggeri2016}. For the sake of clarity  we chose not to display the following categories returned by the program (Figures \ref{fig:feat_atm} and \ref{fig:feat_natm} in Results) due to redundant information ('ester, all','carbon number'), or lack of correspondence to a functional group ('zeroeth group', 'C=C-C=O in non-aromatic ring',  'aromatic CH', 'alkane CH', 'C=C (non-aromatic)', 'alkene CH', 'nC-OHside-a' and 'carbon number on the acid-side of an amide (asa)'). 

\section{Results}\label{sec:res}
In what follows we describe our similarity analysis of atmospheric molecules and how they compare to other compounds found in public molecular datasets. Our initial focus is on molecular structure and composition, followed by a comparison of molecular fingerprint representations. Subsequently, we illustrate the  implications of our analysis in two central applications for machine learning in atmospheric chemistry: computational property prediction and the analysis of mass spectra.

\subsection{Molecular structure comparison}
\begin{figure}[H]
\includegraphics{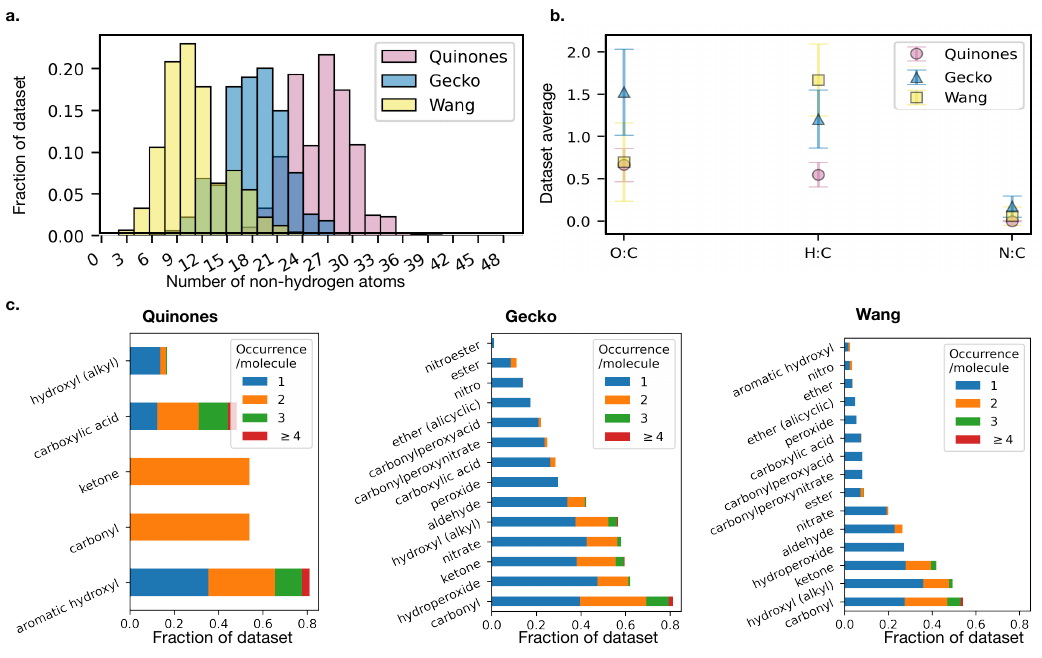}
\caption{Molecular structure analysis of the atmospheric molecules in terms of molecular size as represented by non-hydrogen atom count (a, histogram normalized so bar heights sum to one),  atomic ratios (b), and functional groups (c, present in $\geq$ 10 $\%$ of dataset).}
\label{fig:feat_atm}
\end{figure}

Figure \ref{fig:feat_atm} presents a selected number of molecular features for the three atmospheric datasets included in our work. In Panel a, the molecular size, as measured by the number of non-hydrogen atoms, varies across datasets, averaging to approximately 10, 20, and 30 for the Wang, Gecko, and Quinone datasets, respectively.  The non-hydrogen atoms are mainly oxygen and carbon atoms (see Table \ref{tab:datasets} and Figure \ref{fig:feat_atm}b). The high average O:C ratio of the Gecko molecules suggests that they are appreciably more oxidized than the Wang and Quinone compounds. The Wang molecules are more saturated, as indicated by their high H:C ratio. Functional group analysis reveals common oxygen carrying groups to be hydroxyl, carbonyl, ketone and carboxylic acid groups in all three sets (Panel c). Furthermore, over half of the Gecko molecules contain hydroperoxide and nitrate groups, unlike the Wang (approx. less than a third) and Quinone (absent) compounds. 

\begin{figure}[H]
\includegraphics{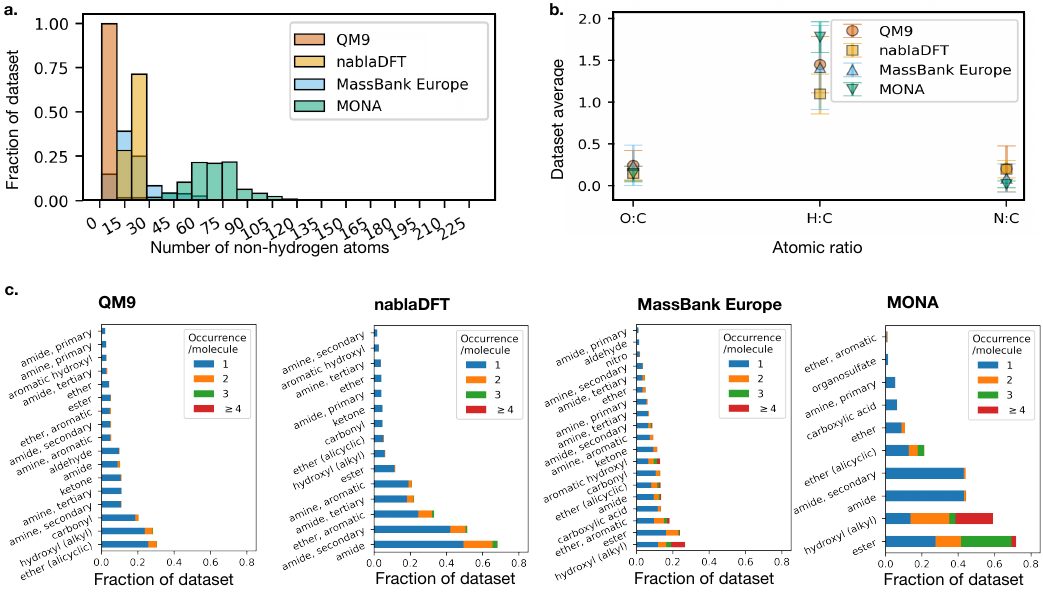}
\caption{Molecular structure analysis of the non-atmospheric molecules in terms of molecular size as represented by non-hydrogen atom count (a, histogram normalized so bar heights sum to one), O:C, N:C, H:C atomic ratios (b), and functional groups (c, present in $\geq$ 10 $\%$ of dataset).}
\label{fig:feat_natm}
\end{figure}

We now turn our focus to the non-atmospheric molecules which primarily comprise metabolites and drug-like compounds. Figure \ref{fig:feat_natm} compares QM9, nablaDFT,  the MassBank Europe and MONA datasets (Table \ref{tab:datasets}). The molecular size in these datasets varies over a wider range than in the atmospheric datasets. While QM9, nablaDFT, and MassBank Europe have a similar size (average 9, 21 and 22, respectively, as measured by the non-hydrogen atom number) to the atmospheric compounds,  the average MONA compounds are larger (68 non-hydrogen atoms). In particular, the largest MONA molecules reach up to 230 non-hydrogen atoms. Such large compounds are not expected to be airborne, except when volatilized for mass spectrometry analysis, or the like.  Compared to the atmospheric molecules, these datasets are markedly less oxidized  and more saturated (low O:C and high H:C, respectively, Panel b). Oxygen-carrying groups like hydroxyls, carbonyls,  esters, and ethers appear in both atmospheric and non-atmospheric datasets (Panel c).  Functional groups like peroxides and nitrates are less prevalent in non-atmospheric than in atmospheric compounds. Finally, amides and amines, the most common nitrogen carrying groups in the non-atmospheric compounds, are rare in our atmospheric datasets. We discuss possible causes for these outlined differences in Section \ref{sec:dis}.

\subsection{Molecular fingerprint similarity}
The molecular structure comparison presented above can be used to identify similarities between atmospheric molecules and other compound classes.  However, in machine learning applications,  the molecules are often represented in a different way,  e.g, using  molecular fingerprints.  Below, we make a similarity comparison using two types of fingerprints: the topological and MACCS fingerprints, to inspect molecular similarities as they would appear to a machine learning algorithm.

\subsubsection{t-SNE clustering}

\begin{figure}[t]
\includegraphics[]{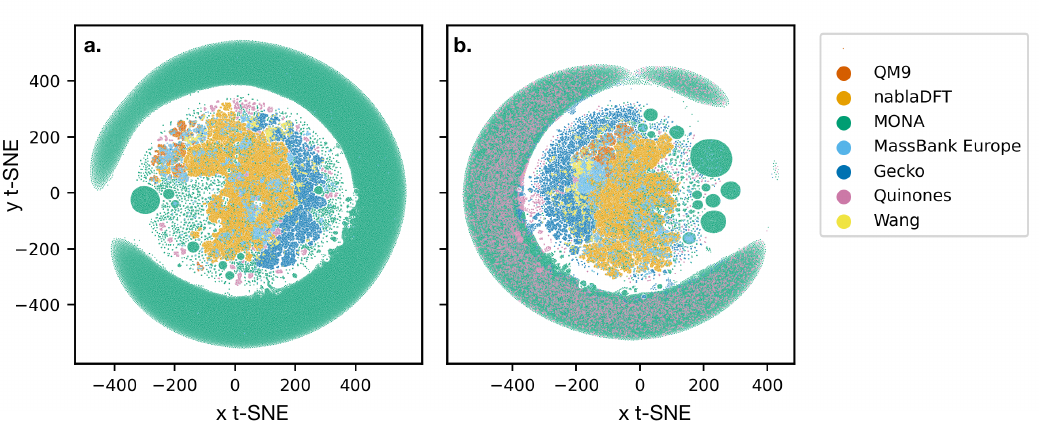}
\caption{Results of t-stochastic neighbor embedding (t-SNE) analysis of different atmospheric and non-atmospheric molecular datasets using a perplexity of 50 and 5000 as the maximum number of iterations. (a) Similarity of topological fingerprints. (b) Similarity of MACCS fingerprints.}
\label{fig:tsne}
\end{figure}

In Figure \ref{fig:tsne}, we compare the atmospheric and non-atmospheric molecules using t-SNE. In t-SNE plots, the degree of similarity among different molecular datasets is discerned by the presence of shared clusters. Figure \ref{fig:tsne}a shows t-SNE clustering using the topological fingerprint as the molecular representation. What stands out is the encompassing cluster or halo of MONA molecules, which does not overlap with molecules from the other datasets. Meanwhile, the nablaDFT dataset forms a central cluster which overlaps with MassBank Europe as well as MONA molecules -- an expected result due to the presence of drug-like molecules in all three datasets. On the left, a group of smaller but similar clusters appears, which also include the QM9 and MONA datasets. Barring a small subset of the Wang and Quinone compounds, these non-atmospheric datasets share no appreciable clusters with the atmospheric compounds which instead form their own, separate clusters. We see both similarities and differences among the atmospheric datasets, as the Gecko and Wang molecules cluster together, but the Quinones form their own clusters. 

In Figure \ref{fig:tsne}b, we have clustered the molecules using the MACCS fingerprint representation. We observe a similar behaviour as for the topological fingerprint. However, the outer ring of MONA molecules now also encompasses portions of the Quinones and Gecko datasets, suggesting that MONA is in part atmospheric-like when viewed through the MACCS representation. In contrast, the topological fingerprint produced more distinct clusters with little overlap between atmospheric and non-atmospheric molecules. 

We tested the robustness of our t-SNE analysis with respect to different perplexity hyperparameter values (Supplementary information, see Figure \ref{sfig:t-sne_top} and Figure \ref{sfig:t-sne_maccs} and refer to Methods for a brief explanation). For perplexities 50 and 100, we find consistent outcomes.  However, for both fingerprints, one central cluster forms at a lower perplexity of 5 encompassing molecules from all datasets. Moreover,  an outer cluster emerges of mainly MONA and Gecko molecules, or a mixture of MONA, Gecko, and Quinones molecules, for the topological fingerprint  (see Figure \ref{sfig:t-sne_top}) and MACCS fingerprint (see Figure \ref{sfig:t-sne_maccs}), respectively.  In summary,  the qualitative t-SNE analysis separates atmospheric and non-atmospheric molecules albeit more so with the topological than the MACCS fingerprint. This finding suggests low similarity across the different chemical domains. 

\subsubsection{Tanimoto similarity distributions}
Next, we conducted a quantitative comparison of molecular fingerprint similarity using the Tanimoto similarity index which ranges from one for perfect to zero for no similarity (see Section \ref{sec:met}). We utilized either Gecko or Wang as our reference atmospheric dataset for two separate comparisons. The compounds in the reference dataset were used to compute pairwise Tanimoto similarities with molecules from other datasets. This analysis was repeated using both the topological and MACCS fingerprints. For facilitating the assessment, we use a high-similarity reference value of 0.4 and a low similarity reference value of 0.1, as detailed in Section \ref{sec:met}.

\begin{figure}[t]
\includegraphics[]{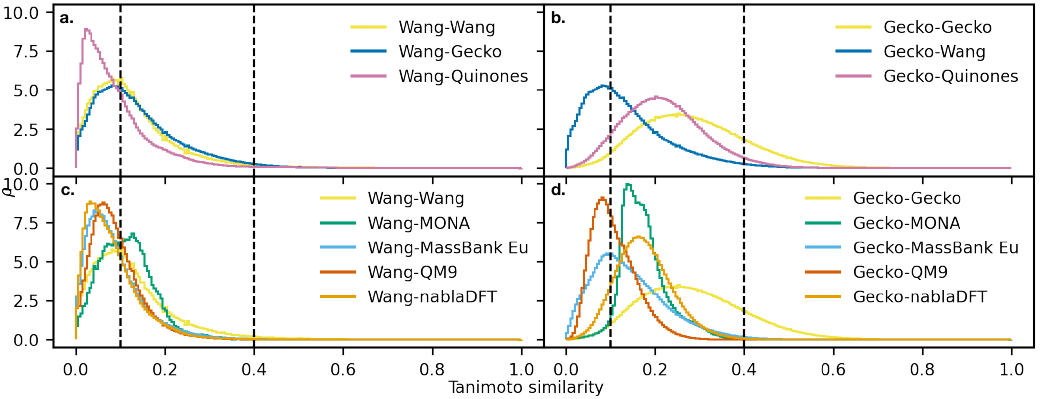}
\caption{The distribution of pairwise Tanimoto similarities between topological fingerprints.  The Tanimoto similarity distribution between atmospheric molecules with the Wang  (a) and Gecko (b) molecules. The Tanimoto similarity between the non-atmospheric molecules with the Wang  (c) and Gecko (d) datasets, respectively. Vertical lines mark our high and low similarities 0.1 and 0.4. The histograms were normalized so that the area under their respective curves integrate to 1.}
\label{fig:tanimoto_wang}
\end{figure}

In Figures \ref{fig:tanimoto_wang}a-d, we analyzed molecular similarity based on the topological fingerprint. The figures depict normalized similarity density distributions, providing insight into the frequency of different similarity values between compounds in the compared datasets. We will utilize the locations of the similarity density distribution peaks to discuss trends.

We begin by establishing a similarity relationship among compounds in our atmospheric datasets.  In Panel \ref{fig:tanimoto_wang}a, the distribution peaks are all below 0.1, indicating that the Wang compounds hold, as a rule, little resemblance to other atmospheric compounds. This is also true amongst the Wang molecules themselves, suggesting that this dataset is diverse.  In Panel \ref{fig:tanimoto_wang}b, we observe that the Gecko dataset is of intermediate similarity (peak between 0.1 and 0.4) to the Quinones set. Moreover, the Gecko molecules have intermediate similarity to each other, and are thus less diverse than in the Wang set. 

Next, we compare the atmospheric and non-atmospheric dataset similarities based on the topological fingerprint in Panels \ref{fig:tanimoto_wang}c and d.  Overall, the Wang compounds have  mostly low similarities to nablaDFT, QM9 and MassBank Europe, albeit with visible fractions of intermediate similarities. Interestingly, MONA is the only dataset with a similarity distribution peak in the region we define as intermediate when compared to the Wang dataset.  Meanwhile, both MONA and nablaDFT have their similarity distribution peaks at intermediate values when compared to Gecko. On the other hand, MassBank Europe and QM9 are on the boundary between low and intermediate similarity values compared to Gecko. Notably, when compared to Gecko, the Wang compounds' similarity distribution peak is at lower values than those of the non-atmospheric datasets  (though the Wang distribution is more right-skewed to intermediate values). In summary, no appreciable amount of similarities of topological fingerprints were in our high similarity region, neither when comparing atmospheric compounds to non-atmospheric molecules, nor among different atmospheric datasets themselves.

\begin{figure}[t]
\includegraphics[]{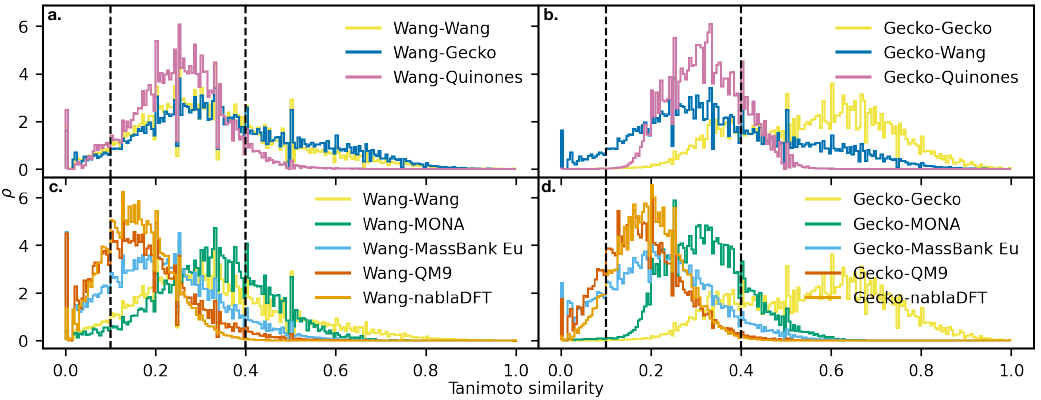}
\caption{The distribution of pairwise Tanimoto similarities between MACCS fingerprints.  The Tanimoto similarity distribution between atmospheric molecules with the Wang  (a) and Gecko (b) molecules. The Tanimoto similarity between the non-atmospheric molecules with the Wang  (c) and Gecko (d), respectively. Vertical lines mark our high and low similarities 0.1 and 0.4. The histograms were normalized so that the area under their respective curves integrate to 1.}
\label{fig:tanimoto}
\end{figure}

An analogous comparison for the MACCS fingerprint (Figure \ref{fig:tanimoto}) revealed similar trends as those for the topological fingerprint. Overall, the similarity distributions are less skewed than those of the topological fingerprint. Moreover, molecules also appear more similar for the MACCS fingerprint. For instance, the Gecko self similarities now peak in the high similarity region (>0.4).  Moreover,  similarity distributions comparing the Quinone and Wang datasets with Gecko compounds peak at intermediate similarity, with visible fractions of the distributions at high similarity values. Also the Wang compounds have appreciably higher similarities to parts of the Quinone compounds. All non-atmospheric similarity distributions peak at intermediate values when compared to both Wang and Gecko (Panels \ref{fig:tanimoto}c-d). In addition, MONA and MassBank Europe have a visible fraction of high-similarity to the Gecko and Wang compounds. We also note that the similarity distributions between atmospheric datasets are broader than between atmospheric and non-atmospheric compounds.

\section{Discussion}\label{sec:dis}
In Figures \ref{fig:feat_atm} and \ref{fig:feat_natm}, we observe a number of features in the molecular structure of atmospheric molecules which set them apart from the other compound classes in terms of functional groups, elemental composition and size (only compared to MONA).  These differences indicate what type of extrapolation a machine learning model would have to do if transferred from one domain to the other.  In particular,  atmospheric oxidation results in a set of organic compounds with distinct atomic ratios and functional groups that are rarely found in other domains. These compounds are primarily made up of carbon, hydrogen, oxygen, and some nitrogen atoms. They stem from volatile emissions, primarily composed of hydrogen and carbon, with nitrogen and oxygen introduced during oxidation. Autoxidation, in particular, increases the oxygen content, resulting in elevated O:C ratios in atmospheric datasets. Oxygen predominantly incorporates into functional groups such as peroxide, nitrate, hydroxyl, carbonyl, and ketones.

Our comparison of nitrogen-containing functional groups instead revealed a lack in amine and amide content in atmospheric compounds compared to the other compound classes. We note that the atmosphere is known to contain numerous reduced nitrogen compounds (estimated to be at least hundreds \citep{Ge2011}). Yet, these compounds are typically presumed to quickly combine with acidic molecules or clusters to form aerosol particles in the atmosphere. Consequently, they are generally excluded from gas-phase oxidation reactions in simulation models such as MCM or Gecko-A, which explains their absence in our study. These artificial biases in the computational generation of atmospheric compounds necessitate scrutiny and awareness when curating atmospheric datasets and developing models based on such datasets depending on application area.

Furthermore, the similarity between molecular representations like fingerprints can unveil whether compounds bear similarity to a machine learning model that utilizes such representations for molecular predictions. Here, the t-SNE and Tanimoto-fingerprint-similarity-metrics revealed low similarities across molecular datasets and compound classes. The t-SNE analysis showed that the atmospheric compounds, besides a certain similarity to MONA and nablaDFT, are distinct as seen through the molecular descriptors ( topological and MACCS fingerprint).  Moreover, the Tanimoto similarity between atmospheric and non-atmospheric molecules is low, and as a rule, below our high-similarity reference value (Figures \ref{fig:tanimoto_wang} and \ref{fig:tanimoto}).  These results reinforce the conclusion that atmospheric compounds should  be considered as out-of-domain for models which have been trained on drug or metabolite like compounds.

Our similarity analysis also revealed that our three atmospheric datasets, albeit sharing molecular features, such as common functional groups and relative atomic ratios, contain a diverse array of compounds. Relative to the comparison of atmospheric and non-atmospheric compounds, we observed that the three atmospheric datasets had a larger fraction of compounds of intermediate similarity. However, we observed few to no high Tanimoto similarity pairs between the three atmospheric datasets for the topological fingerprint, while a larger fraction of high-similarity pairs emerged for the MACCS fingerprint.  These results could be used in future work to curate a diverse set of atmospheric molecules for model training, or assess current blind spots in existing sets. 

Moreover, the Tanimoto similarity analysis of compounds from the same atmospheric datasets (Gecko or Wang) revealed a difference in the degree of  self-similarity which can be traced back to how these datasets were generated.  In Figures \ref{fig:tanimoto_wang} and \ref{fig:tanimoto}, we observed that Gecko molecules exhibit greater similarity to each other, while the Wang compounds are more diverse. This difference in dataset homogeneity can be attributed to the distinct generation processes of the two datasets: Wang was constructed from over 100 precursor compounds, while Gecko from only three. Moreover, in Gecko, the much higher average O:C ratio, and lower H:C ratio, is due to inclusion of more oxidation steps during dataset generation compared to that of Wang. 

The analyses conducted using the t-SNE and Tanimoto metrics reveal varying perspectives on dataset similarity. In the Tanimoto similarity analysis of the atmospheric datasets, the Gecko molecules  have a greater similarity to the Quinone molecules, whereas in the t-SNE analysis, the Wang molecules appear more adjacent. These disparities in perceived similarity arise from the fundamental differences in the algorithms employed by Tanimoto and t-SNE. 

Tanimoto analysis only compares molecular features that are present (represented by ones in the fingerprint) in either molecule, while t-SNE considers both the absence and presence of features (both ones and zeroes) when determining adjacency or similarity in high-dimensional space. Consequently, t-SNE may group molecules based on a common lack of features which the Tanimoto analysis does not. The absence of shared features does not necessarily imply true similarity unless the molecular descriptor captures all molecular structure features, highlighting a limitation of t-SNE for similarity analysis with molecular fingerprints. This distinction in methodology can elucidate why the Gecko and Quinone datasets appear relatively more similar in the Tanimoto analysis compared to the t-SNE analysis, or why the similarity between the Wang and Gecko datasets is relatively high in t-SNE but lower in Tanimoto analysis.

Finally, our comparisons in  Figure \ref{fig:tanimoto_wang} and Figure \ref{fig:tanimoto} highlight the varying degree of atmospheric dataset similarity depending on the molecular descriptor utilized for representing their structures. As alluded to above, a comprehensive molecular similarity measure should be based on an encoding of the entire molecular structure into the descriptor. In this study, we assessed similarity using both topological and MACCS molecular fingerprints. The generally low levels of similarity observed across atmospheric datasets could suggest a potential to develop molecular fingerprints tailored to atmospheric compounds to better capture their unique molecular structure features. Such explorations could be the topic of future work. 

\section{Outlook}\label{sec:out}
Atmospheric compounds constitute a vast and diverse chemical space. Their unique characteristics, coupled with the sheer number of atmospheric compounds, make collecting experimental or high-accuracy computational data both time-consuming and challenging.  Thus, one major challenge to advancing data-driven methods in atmospheric chemistry is the current absence of curated datasets. Therefore, this paper investigated how similar atmospheric molecules are compared to large and openly available datasets that have been utilized in machine learning.  The current  challenges and data gaps elucidated by our similarity analysis and discussion above can be addressed in future work in a number of ways. 

One primary application of our similarity analysis is to serve as a foundation for dataset assembly and to fill data gaps for atmospheric chemistry research. This endeavor will be based on identifying relatively similar datasets from other chemical domains. Once identified,  such molecular datasets can be used to mend data gaps and to improve machine learning development in atmospheric research.   However, since compounds'  similarity is dependent on molecular representation and application area,  data augmentation needs to be judged on a case by case basis.   As more atmospheric research data, whether computational or experimental, becomes available, these models can be further refined.  Below, we list additional considerations when supplementing atmospheric datasets with  larger curated ones from different chemical domains. 

Currently, the limited  number of atmospheric  datasets lack comprehensive coverage of multiple relevant molecular target properties. For example, a dataset may include vapor pressures but not electronic properties or mass spectra. This incomplete coverage leads to data gaps, even when combining multiple datasets, as seen in our investigations.

\begin{figure}[H]
\includegraphics[]{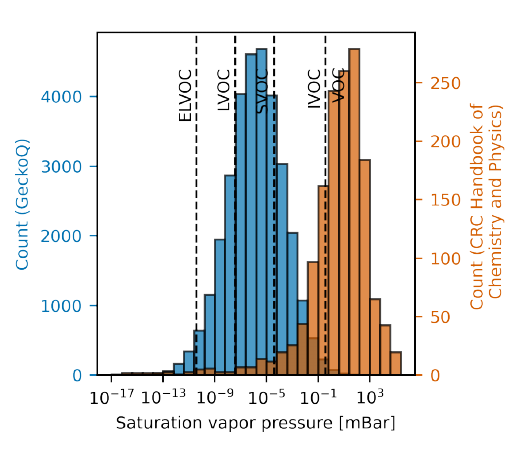}
\caption{Computationally predicted saturation vapor pressure of atmospheric compounds in a subset of the Gecko dataset studied here called GeckoQ  at 298 K (blue), and vapor pressures listed in the CRC Handbook of Chemistry and Physics \citep{Rumble2023} Table entitled 'Vapor Pressure for Inorganic and Organic Substances at Various Temperatures' computed at 298 K using the Clausius-Clapeyron equation. The volatility regions are assigned according to those defined by \citep{Donahue2012}: ELVOC - extremely low volatile organic compounds; LVOC - low volatile organic compounds; SVOC - semi volatile organic compounds; IVOC - intermediate volatile organic compounds; VOC - organic compounds. We assumed the ideal gas law and a molecular weight of an average molecule in organic aerosols (200 g/mol \citep{Donahue2011}).}

\label{fig:feat_psat}
\end{figure}

As already mentioned, to address these cases of missing properties, incorporating existing datasets or models from other disciplines is an alternative to gain larger datasets and potentially improve model training. However,  in addition to structural dissimilarity issues, such data augmentation can be challenging due potential mismatches in target property coverage. For instance, atmospheric particle formation involves compounds with low volatility, which can be characterized by properties like extremely low vapor pressures. These properties often deviate from typical chemistry contexts. In  Figure \ref{fig:feat_psat}, we have compared vapor pressure distributions between oxidized atmospheric compounds from a subset of the Gecko dataset and tabulated values from the Handbook of Chemistry and Physics. This comparison illustrates the underrepresentation of low-pressure target values in generic reference datasets.

Moreover, assessing not only the overlap of target values, but also to carefully examining the target data type is crucial. Each property must be evaluated in the context of its relevance to atmospheric chemistry and its potential impact on the overall dataset integration process. Such considerations become particularly relevant  in mass spectrometry applications. In this study, we compared atmospheric compounds to those found in large mass spectrometric databanks. This choice was based on the central role of mass spectrometry in atmospheric chemistry for studying molecular-level processes \citep{Noziere2015}.

In this study, we have found certain overlaps in terms of molecular fingerprint similarity between atmospheric compounds and molecules in the MassBanks (MONA in particular). However, the fragmentation mass spectrometric techniques commonly employed when generating MassBank data diverge from the prevailing methods utilized in  field campaigns, which predominantly rely on chemical ionization \citep{Noziere2015, sandstroem2024}. Thus, future development of machine learning tools could be directed towards analysis of the mass spectra primarily collected in atmospheric field studies.

A second application of our similarity analysis is for curating new atmospheric molecular datasets.  The persistent challenges to collecting experimental data for atmospheric molecules suggest  that this function primarily fits as a tool for computational studies. Here, the similarity analysis could be used to characterize atmospheric compounds into different types based on their location in chemical space (as defined either by the molecular features or fingerprint, or both). As mentioned in Section \ref{sec:dis}, such characterization can be used to create tailored datasets for analysis, or to construct data-driven analysis tools based on either diverse or niche groups of compounds.

Finally, our study underscores that focus should be given to initiatives aimed at sharing  atmospheric molecular data in openly accessible repositories.  Examples of such initiatives have recently been developed, such as the Aerosolomics project \citep{Thoma2022}. Still,  improving unambiguous identification of atmospheric compounds requires collection of more relevant reference data. Given the diverse techniques and instruments employed in atmospheric science, standardizing data will likely remain challenging. Thus, effective data sharing should include information on data quality and comprehensive metadata, including instrument versions. This information could be considered during development of general predictive models, by, for example, mitigating the impact of instrument versions on data collection and quality. 

\conclusions  
In this study, we compared atmospheric molecules to compounds commonly used to train machine learning models for molecular applications. Assessing molecular structure similarity provides a straightforward means to determine whether atmospheric compounds fall within the scope of existing machine learning methods. This assessment aids in directing the development of machine learning techniques within this relatively unexplored chemical domain. Here, we focused on comparing molecules with two molecular descriptors, the MACCS and topological fingerprints. Analysis of both representations revealed low similarity between progressively oxygenated atmospheric reaction products and non-atmospheric molecules made up of primarily metabolites and drug-like compounds. Notably, the MONA mass spectral library exhibited the highest similarity to atmospheric compounds. Yet, upon scrutiny of molecular size, atomic ratios and common functional groups, we observed disparities between MONA molecules (and other non-atmospheric datasets)  and atmospheric compounds. These discrepancies highlight the need for careful testing and validation before using models trained on MassBank like datasets in atmospheric chemistry.   The differences we observe between chemical domains and between the atmospheric datasets can be used to guide future dataset curation for atmospheric molecular research. Such datasets have laid the foundation for data-driven method development in other chemical domains. Thus, we hope this study will motivate the broader atmospheric chemistry community to establish and contribute to infrastructure for public data-sharing. Closing the current data gap regarding atmospheric compounds will expedite the shift towards a data-driven era in molecular atmospheric research. This advancement will facilitate the development of high-accuracy and high-throughput analysis tools, offering essential insights into the molecular-level atmospheric processes that influence both climate and air quality.

 \codedataavailability{The datasets used for this analysis are all freely available from original publications or the database website (see Table \ref{tab:datasets}). Code and workflow used for data analysis are freely available at this GitLab repository ( \url{https://gitlab.com/cest-group/atmospheric_compound_similarity_analysis}).}

\appendix

\section{t-SNE analysis at different perplexity values}    
\appendixfigures  
\begin{figure}[H]
\includegraphics{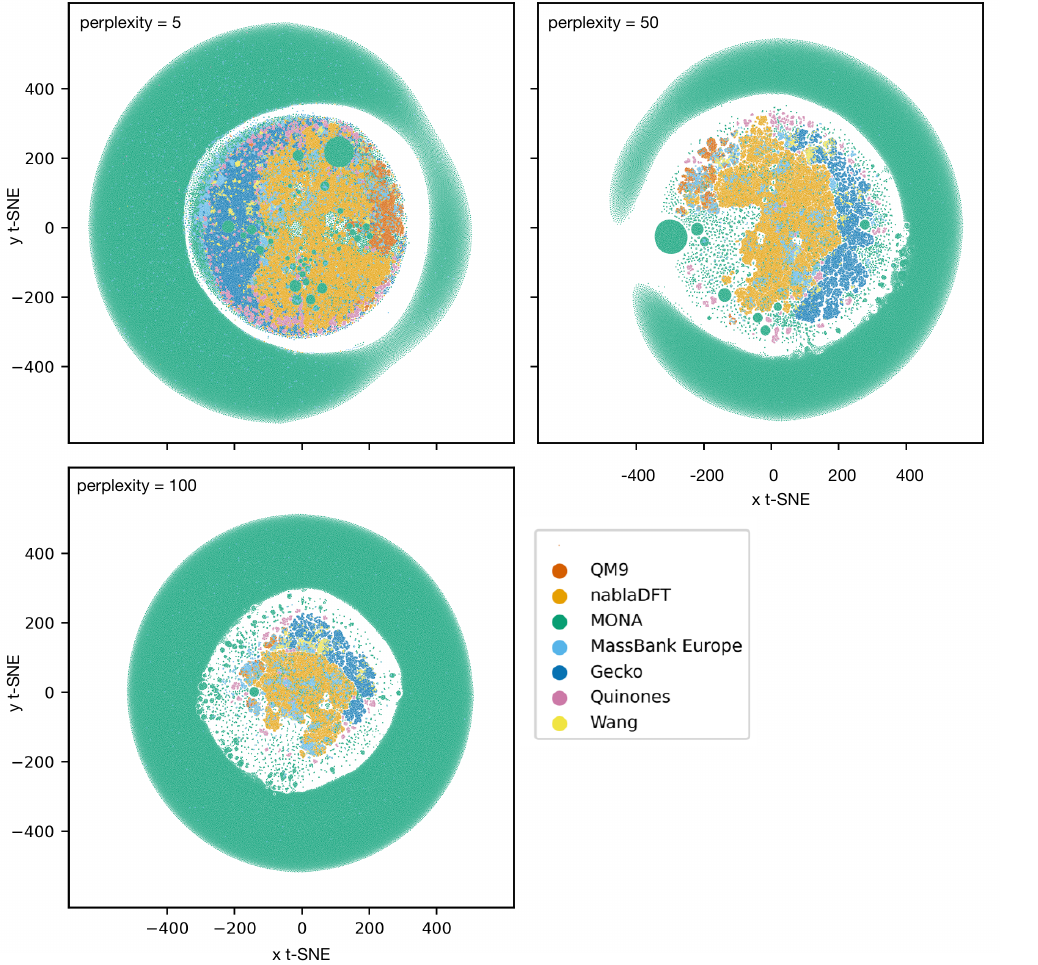}
\caption{(a)The t-SNE analysis of the datasets' topological fingerprints at perplexity values of  5 (top left), 50 (top right), 100 (bottom left).}
\label{sfig:t-sne_top}
\end{figure}
\begin{figure}[H]
\includegraphics{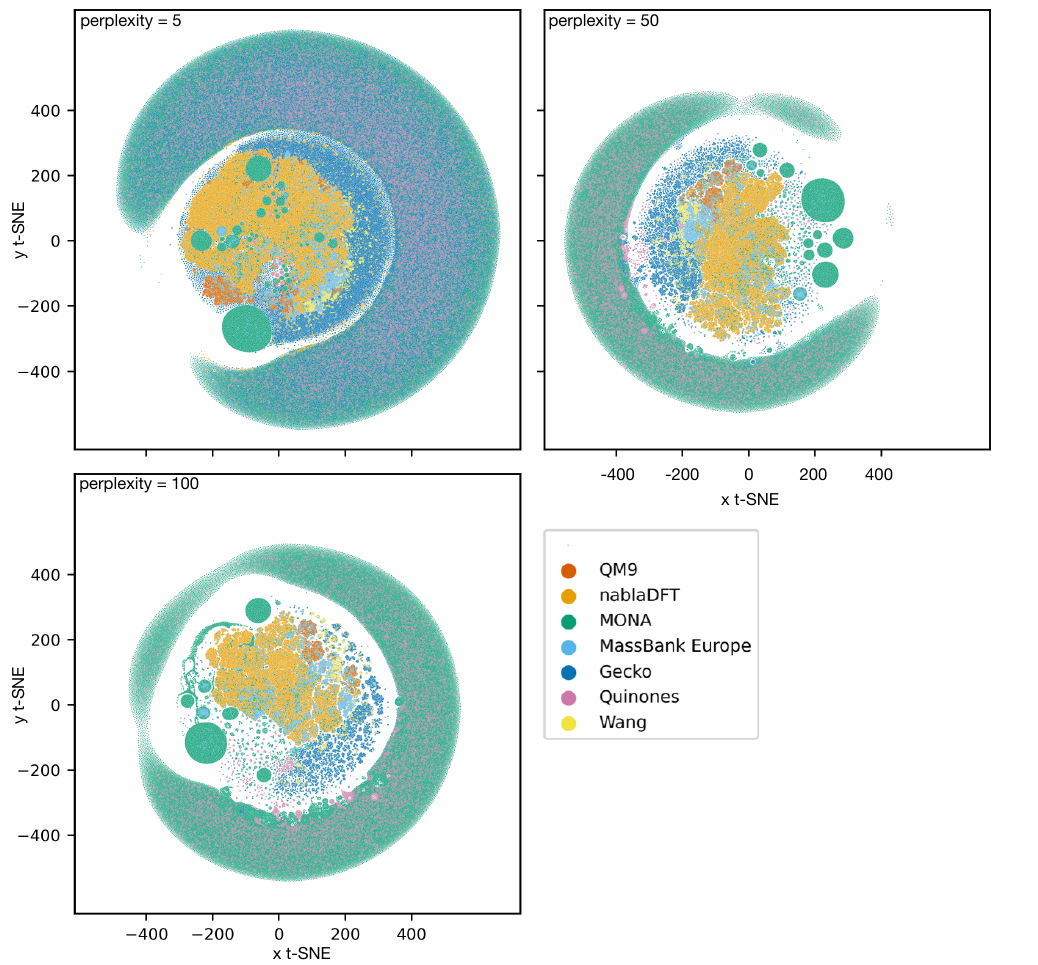}
\caption{ The t-SNE analysis of the datasets' MACCS fingerprints at perplexity values 5 (top left), 50 (top right), 100 (bottom left).}
\label{sfig:t-sne_maccs}
\end{figure}

\noappendix       





\authorcontribution{H.S. - Conceptualization, Investigation, Data Curation, Formal analysis, Visualization, Validation. Writing - Original draft. P.R - Conceptualization,  Writing - Original draft, Supervision, Funding acquisition.} 

\competinginterests{The authors declare no competing interests.} 


\begin{acknowledgements}
The authors wish to acknowledge T. Kurten and M. Rissanen for insightful discussions. This study was supported by the Academy of Finland through Project No. 346377 and the EU COST Actions CA18234 and CA22154. We further acknowledge CSC-IT Center for Science, Finland and the Aalto Science-IT project.
\end{acknowledgements}






\bibliographystyle{copernicus}
\bibliography{manuscript.bib}

\end{document}